# Oxide MBE – ARPES at SSRL Beamline 5-2

MAKOTO HASHIMOTO,[1] YONG ZHONG,[2,3,4,5], AND DONGHUI LU[1]

[1]*Stanford Synchrotron Radiation Lightsource, SLAC National Accelerator Laboratory, Menlo Park, California, USA*
[2]*Stanford Institute for Materials and Energy Sciences, SLAC National Accelerator Laboratory, Menlo Park, California, USA*
[3]*Advanced Light Source, Lawrence Berkeley National Laboratory, Berkeley, California, USA*
[4]*Geballe Laboratory for Advanced Materials, Stanford University, Stanford, California, USA*
[5]*Department of Applied Physics, Stanford University, Stanford, California, USA*

## Introduction

Quantum many-body materials display a wide range of emergent phenomena and quantum phases due to the complex interplay of charge, spin, orbital, and lattice degrees of freedom [1, 2]. A few examples are strong correlation-driven metal-to-insulator transition, charge density wave, high-$T_c$ superconductivity, and various topological phases. These materials are of great interest for condensed matter physics and future applications. To gain a deeper understanding of such emergent phenomena, it is crucial to characterize the electronic states in detail. Over the past two decades, angle-resolved photoemission spectroscopy (ARPES) has become a standard experimental tool to characterize the electronic structure of quantum materials [3, 4]. ARPES measures electrons in single-crystalline solids with energy and momentum resolution, realized through the measurement of the kinetic energies and emission angles of photoelectrons. It provides information relevant to the occupied part of the single-particle spectral function. Visualization of the band dispersion allows direct comparison with theoretically calculated band dispersions. Detailed investigations of the electronic states reveal important information on electron-electron correlation, electron-boson couplings, and energy gaps in the single-particle excitation, which often lead to the discovery of physics beyond theoretical predictions.

When combined with synchrotron light sources, ARPES becomes an even more powerful experimental tool [3, 4]. Synchrotron light sources provide continuous photon energy tunability over a wide energy range. By varying the photon energy, the $k_z$ dependence of the three-dimensional electronic structure can be mapped out. On the contrary, surface electronic states are two-dimensional, so the absence of $k_z$ dependence is a key signature for disentangling it from three-dimensional (3D) bulk electronic states. In addition, varying the photon energy from vacuum ultraviolet to soft X-ray changes the photo electron escape depth from more surface-sensitive <1 nm range to more bulk-sensitive up to a few nm. Meanwhile, the photon energy can be optimized to empirically enhance the feature of interest. When the photon energy is tuned to a resonance edge, corresponding orbitals can be substantially enhanced. Furthermore, modern beamlines with full polarization control allow us to disentangle the orbital character of the measured bands in detail. As such, high-flux and high-resolution beamlines coupled with modern high-resolution and high-efficiency electron spectrometers are popular tools for comprehensive electronic structure characterization.

Synchrotron-based ARPES has proven to be a powerful technique for investigating different types of quantum materials. These materials can be broadly classified into three groups. The first group includes single crystals, which are typically measured by cleaving or fracturing the sample under an ultra-high vacuum to obtain the clean surface required for ARPES. This is the most conventional approach and provides robust and reliable data when high-quality single crystals are available and can be easily cleaved. The second group includes exfoliated 2D materials and their heterostructures, which have become a central topic in the field over the last decade. Van der Waals single crystals can be thinned down to the atomically thin limit, and in addition, they can be stacked with twisted angles. Although the size of exfoliated thin crystals is typically small (<10 microns), improved synchrotron beam spot size with advanced optics has opened up a new area of research into 2D materials using ARPES. The third group includes thin films grown by techniques such as molecular beam epitaxy (MBE) and pulsed laser deposition (PLD). Thin film growth enables more flexible material design and faster feedback. It is also possible to grow single crystalline thin films with tailored properties that are not available as bulk crystals. Moreover, systematic measurements unique to thin films, such as substrate, termination, and thickness dependence, as well as heterostructure are possible. These thin film crystals are typically large enough for ARPES measurements. As a result, ARPES on thin films has become a standard approach for electronic structure characterization using both synchrotron and lab-based photon sources.

However, ARPES on thin films sometimes presents challenges as the sample surface needs to be clean enough for ARPES. Some thin film sample surfaces are stable enough for ex-situ measurements, often with a capping layer and a de-capping/post-annealing process prior to measurements. For thin films that cannot undergo these processes, vacuum sample suitcases may be used to preserve the cleanliness of the sample's surface. This option is available at some synchrotron ARPES facilities. To obtain the best possible sample surface cleanliness, thin film growth chambers and ARPES chambers need to be connected for *in-situ* measurements. This allows for quick transfer of the samples under ultrahigh vacuum from the growth chamber to the measurement chamber ARPES systems with *in-situ* thin film growth capabilities are becoming increasingly popular for studying quantum materials by design, both at synchrotrons and in laboratories. This *in-situ* growth approach is particularly crucial for oxide thin films, as they are more vulnerable to air and other gases. Several laboratories have built sophisticated systems that integrate ARPES and oxide MBE [5–7]. However, synchrotron-based ARPES setups coupled with *in-situ* oxide MBE are rare [8], even though such a combination is expected to open exciting opportunities in quantum materials research.

In this article, we highlight our synchrotron ARPES studies of oxide thin films grown by *in-situ* connected MBE at beamline 5-2 of Stanford Synchrotron Radiation Lightsource (SSRL). This combination makes it possible to investigate various complex oxides that are not possible otherwise. After introducing the beamline/endstation capabilities, we will review the oxide MBE system and our two recent case studies, quasi-1D chain copper oxides [9] and La-based cuprates [10], to demonstrate the powerfulness of the oxide MBE-ARPES technique.

## SSRL ARPES beamline

The Stanford Synchrotron Radiation Lightsource is a $3^{rd}$ generation synchrotron facility with a 3 GeV storage ring [11]. Since the 1980s, SSRL has been at the forefront of developing synchrotron-based ARPES. Beamline 5 is a dedicated ARPES beamline that currently features an EPU (elliptically polarizing undulator), which hosts two ARPES branches, namely beamline 5-2 [12] and beamline 5-4 [13].

Beamline 5-4 is a branch equipped with a Normal Incidence Monochromator (NIM) that covers a photon energy range of 7–40 eV [13]. It has a resolving power of ~20,000 and a photon flux greater than $10^{11}$ photons/s at 10,000 resolving power. Coupled with a SCIENTA R4000 electron analyzer, high-energy resolution (~3 meV) measurements can be performed with high stability (<1 meV/day). The endstation is equipped with a home-built 5-axis low-temperature manipulator that operates at 6–400 K. The manipulator also has a ceramic heater mounted on the sample stage for local sample heating. This feature allows us to suppress outgassing during temperature-dependent measurements thus maintaining an ultra-high vacuum of better than $3 \times 10^{-11}$ Torr. This minimizes sample surface degradation, which is most strikingly demonstrated by high-precision measurements of cuprate superconductors $Bi_2Sr_2CaCu_2O_{8+\delta}$ [14, 15]. Most of the ARPES studies performed at beamline 5-4 are on single crystals of correlated electron systems [16, 17]. However, a vacuum suitcase or decapping of the capping layer has been used for thin film studies at beamline 5-4, as demonstrated in earlier synchrotron-ARPES studies of FeSe films [18–20], for example.

Beamline 5-2, which has been operational for general users since 2017, is a newer branch featuring a plane grating monochromator (PGM) [12]. Figure 1 shows the optical layout of the beamline, which uses three variable line space plane gratings to cover a photon energy range of 18–200 eV. The ultimate resolving power of the beamline is 40,000, with a photon flux of >$10^{12}$ photons/s at a resolving power of 10,000. The last elliptical-toroid mirror is the single refocusing optics, which focuses the beam down to a spot size of 30 x 6 µm$^2$ at the sample measurement position, much smaller than the one at beamline 5-4 (100×50 µm$^2$). The photon energy dependence of the beam position and spot size has been carefully calibrated and optimized, resulting in a consistent beam position within a few µm and optimal beam focus over the entire photon energy range.

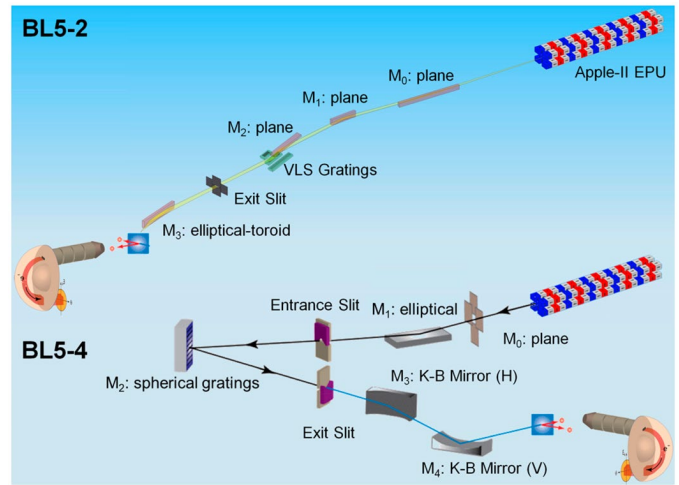

*Figure 1: Schematic of optical layout of SSRL Beam Line 5. Two complementary high-resolution ARPES branchlines for quantum materials, BL5-2 and BL5-4, share the Apple-II type EPU. While the NIM branchline, BL5-4, covers 7–40 eV photon energies, the PGM branchline, BL5-2 covers 18–200 eV photon energies. The BL5-2 ARPES endstation is in-situ connected to a suite of thin-film growth/characterization chambers.*

Figure 2A shows a CAD rendering of the BL5-2 endstation, which comprises a two-level measurement chamber (upper and lower chambers), a sample preparation chamber with an e-beam heating stage and sample garage, and a sample load-lock. Additionally, a UHV sample highway (Veeco) connects the ARPES system *in situ* to thin-film growth chambers, which will be detailed in the next section. Furthermore, the preparation chamber has an open port with a buffer chamber for docking sample suitcases brought in by users. This enables in-vacuum sample transfer of thin films and thin flakes prepared elsewhere, ensuring a higher sample surface quality.

The BL5-2 endstation is equipped with a SCIENTA DA30-L electron analyzer. The total energy resolution of < 3 meV has been demonstrated in high-resolution gap measurements on real samples owing to the high energy resolution of the analyzer and beamline. The analyzer features advanced electron deflectors that can cover an angular range of a 30° cone, allowing for full Fermi surface mapping without the need to rotate the manipulator. The combination of small beam spot size and deflector mapping make it possible to collect high-quality data from small or inhomogeneous cleaved samples, as well as relatively large exfoliated thin crystals (over 20 microns). The manipulator is a home-built 6-axis manipulator. The operating temperature range is 7–400 K, which is achieved with a Janis ST-400 cryostat coupled with a recirculating gas cooler (RGC). Similarly to the manipulator for beamline 5-4 endstation, a ceramic heater for local heating is installed. The sample stage is further equipped with four electrodes for a versatile sample environment. Utilizing these electrodes, Piezo-based uniaxial strain device has been demonstrated, and the nematicity in iron-based superconductor [21–23] has been studied. Further development of other sample environments, such as electric current, magnetic field, and electric gating, is ongoing. A microscope mounted on a reentrant viewport in the upper chamber and multiple telescopes in the upper and lower chambers are implemented for the optical characterization of strained samples and thin flakes, and for efficient sample alignment to the beam and analyzer focus, respectively.

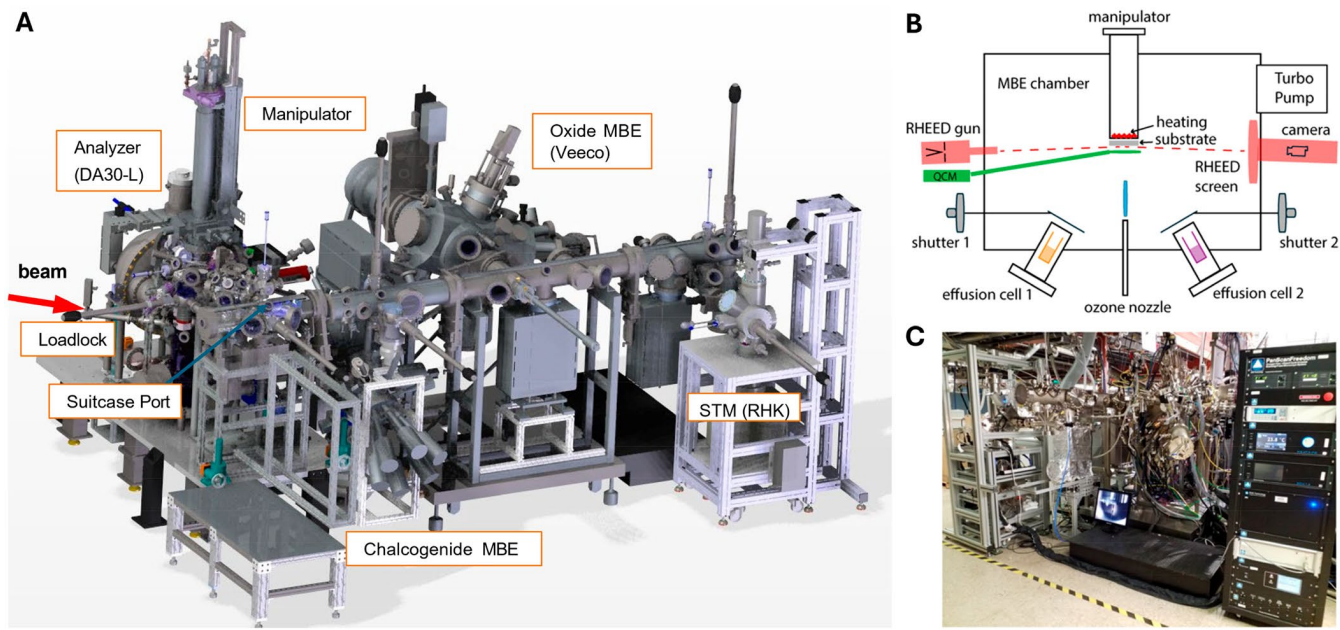

*Figure 2: A. Schematic of the endstation at BL5-2. It consists of an ARPES measurement chamber equipped with a Scienta DA30-L analyzer, sample preparation chamber, loadlock, suitcase port, chalcogenide MBE, oxide MBE, STM, all connected through the UHV transfer system. B. Basic components in an oxide MBE system, including effusion cells, RHEED, QCM, purified ozone gas delivery system, sample manipulator, and pumps. C. Front view of a Veeco Gen930 oxide MBE system.*

ARPES measurements are performed through a Python-based home-developed data acquisition software [24]. Under this Python program, PEAK servers for SCIENTA analyzer control, a SPEC server for beamline control, and EPICS servers for various controls are running background. A screenshot of the user interface is shown in Figure 3. It consists of both graphical user interfaces and script-based user interfaces for automated measurements. This is fully integrated software for efficient, flexible, and well-controlled measurements yet easy to use for non-expert users. Further development of measurement software and standard data analysis tools is ongoing. As such, the ARPES beamline achieves high performance in many aspects. In the next section, we focus on a unique capability of this beamline, which is the in-situ connected oxide thin film growth.

## Oxide MBE

A unique feature at SSRL beamline 5-2 is that the ARPES endstation is *in-situ* connected to thin-film growth facilities: an oxide MBE chamber and a chalcogenide MBE chamber, which enable us to establish a joint synthesis-characterization platform to investigate the electronic structures of epitaxial quantum materials with tailored properties and unprecedented accuracy. Due to the space limitation, here we focus on the oxide MBE system and two related research topics.

Compared with other epitaxial techniques, MBE can control the material synthesis with atomic level precision, paving a new avenue for tailoring the physical properties and manipulating the emergent phases. There are five essential ingredients for a modern oxide-MBE system, as shown by the schematic diagram in Figure 2B. First, an ultrahigh vacuum (UHV) environment can ensure low defect concentrations during the synthesis process. In practice, a series of mechanical and turbo pumps are employed to reach the UHV condition. Second, precise control of the beam flux is crucial to growing complex oxide compounds. Effusion cells with PID control can evaporate a stable molecular or atom flux via temperature tuning. Moreover, a quartz crystal monitor (QCM) is often used to calibrate the beam flux before the actual film growth. Third, reflection high-energy electron diffraction (RHEED) is a powerful technique to realize a real-time monitor of the film synthesis. In particular, the RHEED oscillation reflects the growth rate of a single unit cell. Fourth, the oxygen/ozone partial pressure is a key parameter in the synthesis of transition-metal oxides. Distilled ozone can provide efficient oxidization capability, which has been widely used in the growth of cuprate superconductors and other strongly correlated oxides. Lastly, automatic control of the shutter on/off sequence can facilitate the sample synthesis in an atomic layer-by-layer mode, which is the unique advantage of designing metastable structures and artificial compounds by exploiting the MBE method.

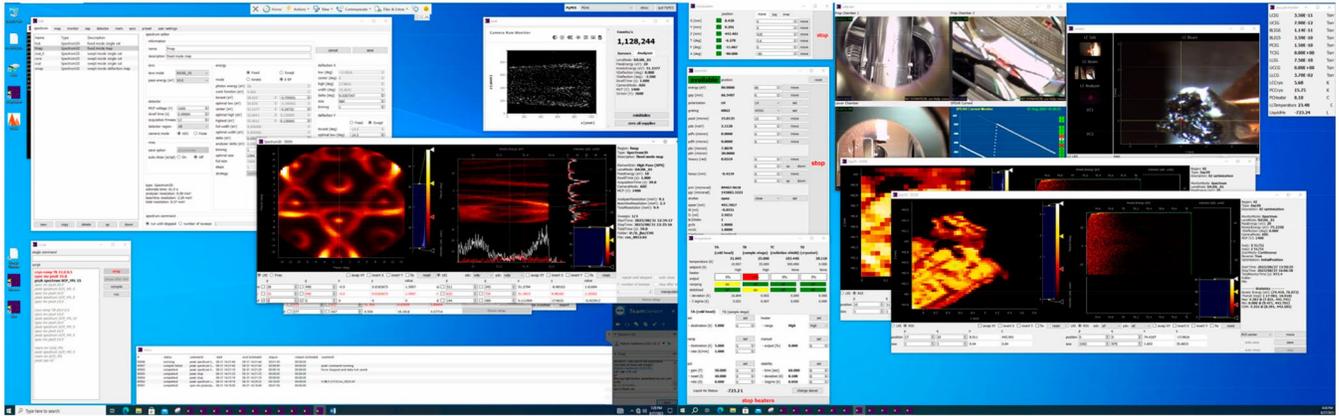

*Figure 3: A screen shot of the user interface of the data acquisition system. It consists of intuitive graphical user interfaces and script-based user interface for more flexible automated measurements. Under this python-based program, PEAK servers for the analyzer control, spec server for beam line control, and epics servers for various controls such as the sample manipulator, are running.*

In BL5-2, a Veeco Gen930 oxide-MBE chamber is *in-situ* connected to the ARPES endstation through an in-vacuum sample transfer system. The MBE chamber is equipped with eight effusion cells, an Inficon SQM-160 QCM, a Staib RHEED, and an ozone generation instrument from DCA (Figure 2C). The base pressure is below $2 \times 10^{-10}$ Torr. The Molly software program allows us to write recipes for controlling the shutter on/off sequence and duration, providing great flexibility in practical growth activities. Furthermore, there is a customized trolly setup *in-situ* transferring the thin films from the growth chamber to the ARPES endstation. The base pressure is better than $1 \times 10^{-10}$ Torr during the transfer process. By integrating the advantages of the epitaxial technique and advanced spectroscopy, we will illustrate the power of this joint platform in the following two examples.

### Benchmarking the microscopic model in 1D cuprates

Although three decades have passed since the milestone discovery of high-$T_C$ superconductivity in cuprates, the microscopic understanding of the pairing mechanism remains an open question in condensed matter physics. One big obstacle is that many-body Hamiltonians in two or more dimensions have no exact solutions under the current paradigm of theoretical and numerical techniques, making a quantitative comparison between theory and experiment a big challenge. Instead, a one-dimensional (1D) system can be solved exactly in the microscopic Hamiltonian like the Hubbard model, providing a promising route to improve the understanding of many-body physics in strongly correlated materials. In a 1D system, strong correlation fractionalizes the electron or hole into spinon and holon with different velocities [25]. Such spin-charge separation has been confirmed by ARPES measurements in undoped 1D cuprate [26]. However, the lack of information on doped carriers makes it difficult to have a comprehensive understanding of the microscopic model in the 1D system. In particular, it remains undetermined whether the Hubbard model contains all of the essential ingredients for describing the general properties of cuprates. Therefore, a doping-dependent study of 1D cuprate is imperative to clarify the above puzzle.

We synthesized the 1D $Ba_{2-x}Sr_xCuO_{3+\delta}$ (BSCO) thin films with a wide range of hole doping by exploiting the oxide-MBE technique. The doping level of BSCO is determined by the number of interstitial oxygen $\delta$, which can be tuned by *in-situ* ozone and vacuum annealing processes. The chain-type BSCO forms twinned domains on $SrTiO_3$ substrate with different orientations. Figure 4A displays the doping-dependent Fermi surface maps on BSCO thin films. The two perpendicular Fermi surfaces originate from the twinned domains. Figure 4B and C show the energy-momentum cuts and corresponding second derivatives along the Brillouin zone boundary, in which the spin-charge separation is well identified: (i) The holon branch (blue dashed line with "h") crosses the Fermi level at the Fermi momentum $k_F$; (ii) The spinon branch (red dashed line with "s") flattens near $k_1 = 0$, and merges into the holon branch for higher momentum. These features are consistent with previous observations in undoped 1D cuprates. In addition to the spin-charge separation, an additional folded branch, denoted by "hf" in Figure 4C, is clearly observed in lightly doped BSCO and fades quickly at higher doping, as shown more directly in the momentum distribution curves for 9%, 14%, and 33% dopings (Figure 5A–C). In comparison with the theory of the single-band Hubbard model (Figure 5D and E), the hf branch arises from a holon-holon interaction mediated by the spin super-exchange [27].

Despite a consistency in the basic dispersion, we find that the simple Hubbard model is deficient in addressing additional spectral features. For example, the Hubbard model predicts the presence of another bent branch around $2\pi$-$3k_F$ (called $3k_F$ branch), as well as the relatively stronger intensity of the $3k_F$ branch by comparing to the hf feature (Figure 5D). However, the experimental observation is the opposite. This implies that a simple Hubbard model with strong on-site repulsive U is insufficient to describe the holon-holon interaction in 1D cuprates. Instead, a sizeable attractive near-neighbor interaction V must be added into the Hubbard Hamiltonian to reproduce the experimental spectra. As shown in Figure 5F and G, this attractive V enhances the spectral weight of the hf branch and suppresses the $3k_F$ branch, as well as qualitatively tracks the doping-dependent evolution of the hf feature. The pure interaction between electrons should be repulsive, so the only possible origin of this effective attraction would be coupling to some bosonic excitations. Here, we postulate phonon as a candidate in light of the evidence of electron-phonon coupling in a variety of cuprates [28, 29]. Considering structural and quantum chemistry similarities among cuprates, this attraction may play an important role in high-$T_C$ cuprate superconductors.

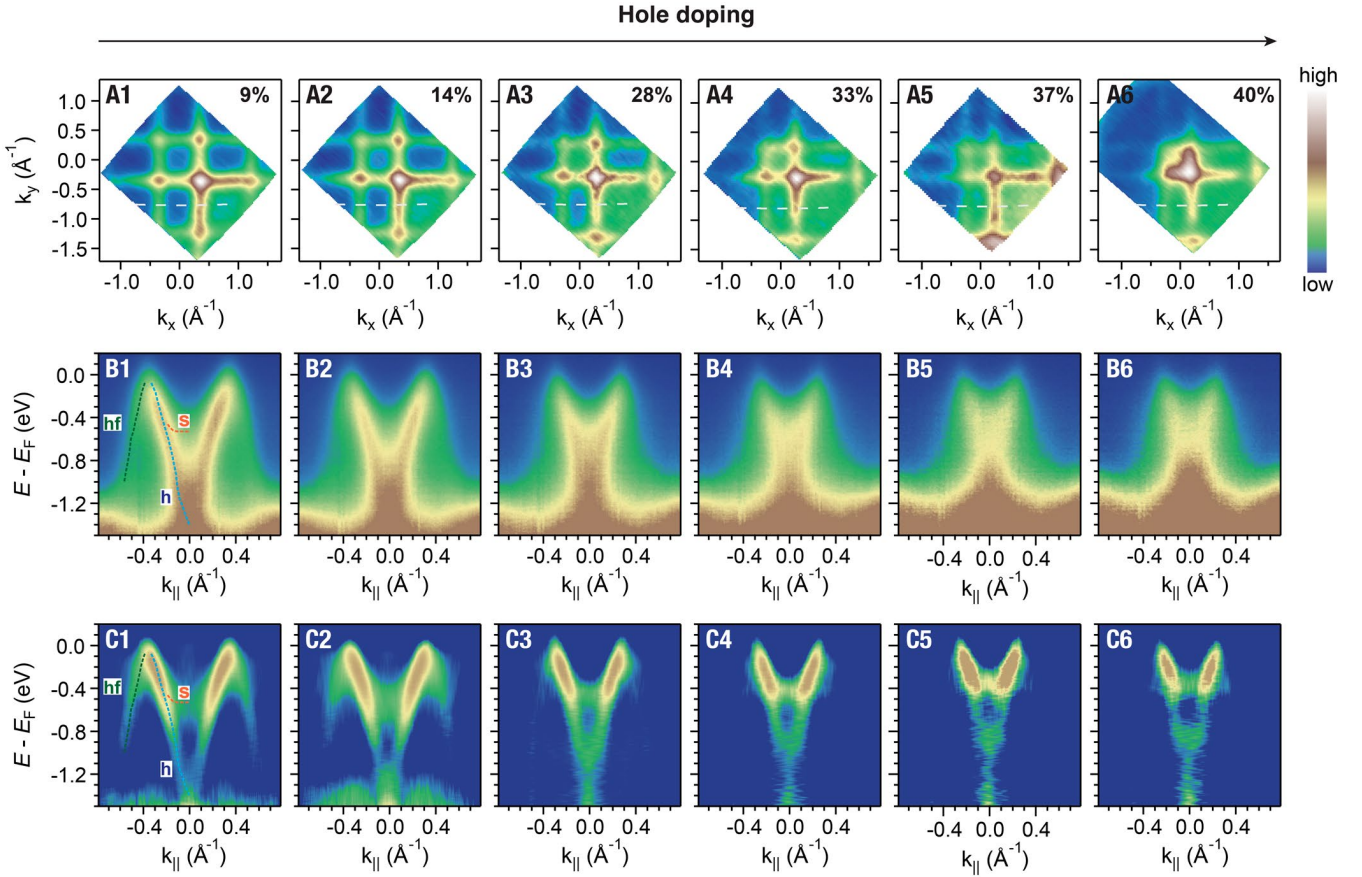

*Figure 4:  Doping-dependent ARPES spectra of 1D cuprate BSCO. (A1 to A6) Fermi surface maps with different doping levels. (B and C) Doping-dependent ARPES spectra and the corresponding second derivatives taken along the white dashed cuts in (A1) to (A6). Dashed blue, red and green lines denote the holon, spinon and holon folding features.*

### Lifshitz transition in high-$T_C$ cuprate superconductors

Physics underlying the rich electronic phase diagram in hole-doped cuprates remains an active field of research. With the change of doping, a Lifshitz transition (LT) of the Fermi surface is expected when the chemical potential crosses the van Hove singularity (VHS), at which the Fermi surface transforms from a hole-like pocket around $(\pi, \pi)$ to an electron-like pocket around $(0, 0)$. Recent experimental progress, including the putative termination points of pseudogap [30], the presumed quantum critical points seen in electronic specific heat and Hall number [31], and the anomalously low superfluid density in electromagnetic response in the over-doped regime [32], highlights the richness of the electronic phase diagram. On the theoretical front, the presence of VHS was suggested to be important for the enhanced $T_C$ and phase fluctuations [33]. Two-dimensional Hubbard model simulations proposed that the LT transition has a strong influence on the phase diagram of hole-doped cuprates [34]. Given the complexity of experimental phenomenology and theoretical postulations, a systematic study of the electronic structure across the LT is desirable. LTs have been explored in many families of cuprates by ARPES measurements, such as La$_{2-x}$Sr$_x$CuO$_4$ (LSCO) [35], YBa$_2$Cu$_3$O$_{6+\delta}$ [36], (Bi,Pb)$_2$Sr$_2$CuO$_{6+\delta}$ [37] and Bi$_2$Sr$_2$CaCu$_2$O$_{8+\delta}$ [38]. Among them, LSCO is an ideal platform because the critical points in other cuprate families are usually close to the materials' doping solubility limit. Nonetheless, the three-dimensionality in LSCO broadens the originally critical point to a finite doping range, within which "electron-like" and "hole-like" features coexist and manifest at different $k_z$. Therefore, a detailed doping-dependent and $k_z$-dependent photoemission study is needed to dissect the LT in LSCO.

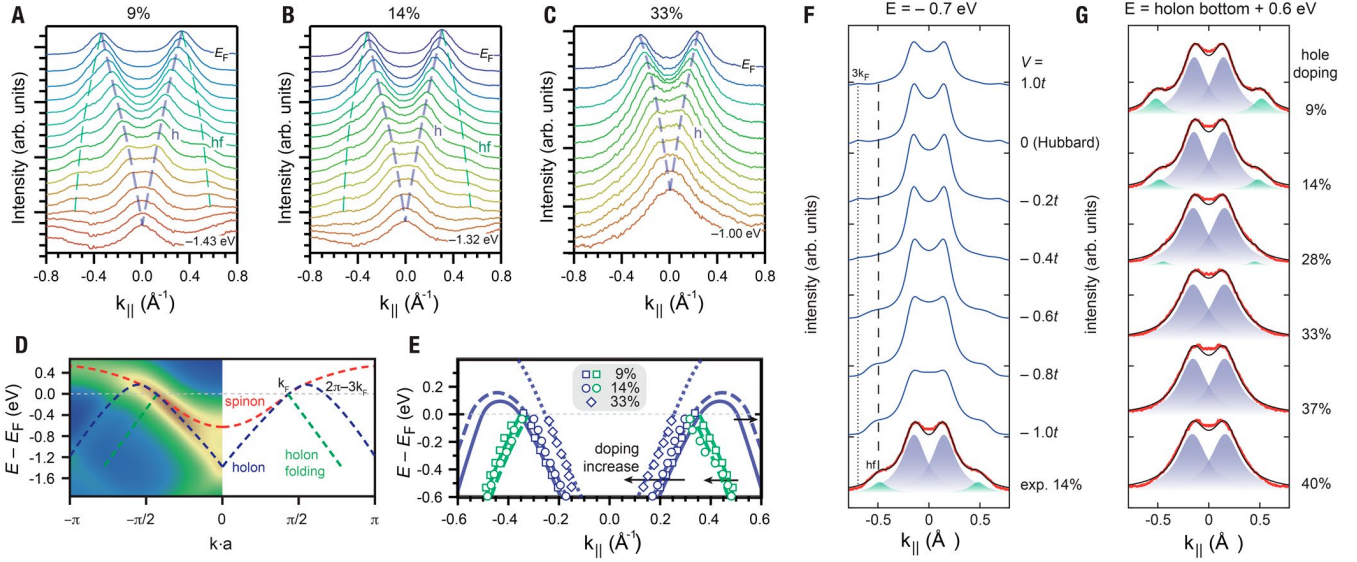

*Figure 5: Holon folding and the attractive interaction in 1D cuprate BSCO. (A to C) Momentum distribution curves (MDCs) at 9%, 14%, and 33% hole dopings. Purple and green dashed lines indicate the holon and holon folding branches. (D) Single-band Hubbard model simulation of the spectra at 12.5% doping level. Note that the holon folding feature is relatively weak in the Hubbard model simulation without additional terms. (E) MDC-derived dispersions for different doping levels. (F) Comparison between the simulation with a near-neighbor interaction term V (blue curves) and experimental MDC at −0.7 eV for 14% doping (red curve). Black curve is a fit with two Lorentzian peaks shown in blue (major) and green (hf). (G) Doping-dependent experimental MDCs at 0.6 eV above the holon bottom.*

Here, we prepared the LSCO thin films with a wide range of doping levels. Then we performed systematic photon energy dependent ARPES measurements with improved spectral quality due to the ultraclean protection of the sample surface. Three-dimensional Fermi surface can be probed with in-plane momentum $k_∥$ (a combination of $k_x$ and $k_y$) and out-of-plane momentum $k_z$. Firstly, as displayed in Figure 6, we focus on the in-plane Fermi surface evolution at a selected photon energy $hν = 70$ eV. According to the Luttinger theorem, we calculate the doping levels via counting the filled states enclosed by the Fermi surface contours. There is a clear LT between the $x = 0.17$ and $x = 0.22$ LSCO samples as the Fermi surface evolves from hole-like to electron-like. Secondly, we explore the $k_z$ dependence of the Fermi surface by tuning the photon energies from 60 eV to 170 eV. A periodic $k_∥$-$k_z$ dispersion on the $x = 0.22$ sample is shown in Figure 7A. The tight-binding model is employed to fit the Fermi surface contours. 3D Fermi surface of $x = 0.22$ sample is illustrated in Figure 7C, where the red contour denotes the in-plane Fermi surface at $hν = 70$ eV. To quantify the LT regime accurately, we calculate the doping-dependent antinodal binding energy (Figure 7B), which pins the LT into doping range $x = 0.20$–0.21.

The established three-dimensionality of the Fermi surface is crucial for understanding the specific heat anomaly in LSCO. The LT range is close to the doping of the normal-state electronic specific heat peak in LSCO cuprates. With the band parameters determined by fitting the electronic dispersion, we calculated the doping-dependent electronic specific heat and found a quantitative agreement with the thermodynamic microcalorimetry experiment in the vicinity of LT [39, 40], as shown in Figure 7D. This remarkable consistency between ARPES and thermodynamics indicates that the Fermi surface LT contributes dominantly to the specific heat peak in LSCO. Quantum critical fluctuations, if exist, only have a minor role on this anomaly. To conclude, our study provides a band theory explanation for the specific heat peak in LSCO, without the need to include the substantial contribution from quantum critical fluctuation.

## Summary and outlook

In this paper, we introduced synchrotron-based MBE-ARPES studies at SSRL BL5-2. The PGM branchline 5-2 covers a wide photon energy range with high flux, high resolution, polarization control as well as small beam spot, complementary to the NIM branchline 5-4. The integration of the oxide MBE to such ARPES beamline/endstation 5-2 further opens up opportunities to study oxide thin films not available in the bulk form. Such a combination allows us to study the unique physics of quantum materials in oxide thin films. After introducing the experimental setups at the beamline, we presented a couple of case studies to illustrate the powerfulness of the coupled system, including doped 1D cuprates not available in the form of single crystals [9] and a systematic investigation of LSCO [10] realized by the combined experimental setup.

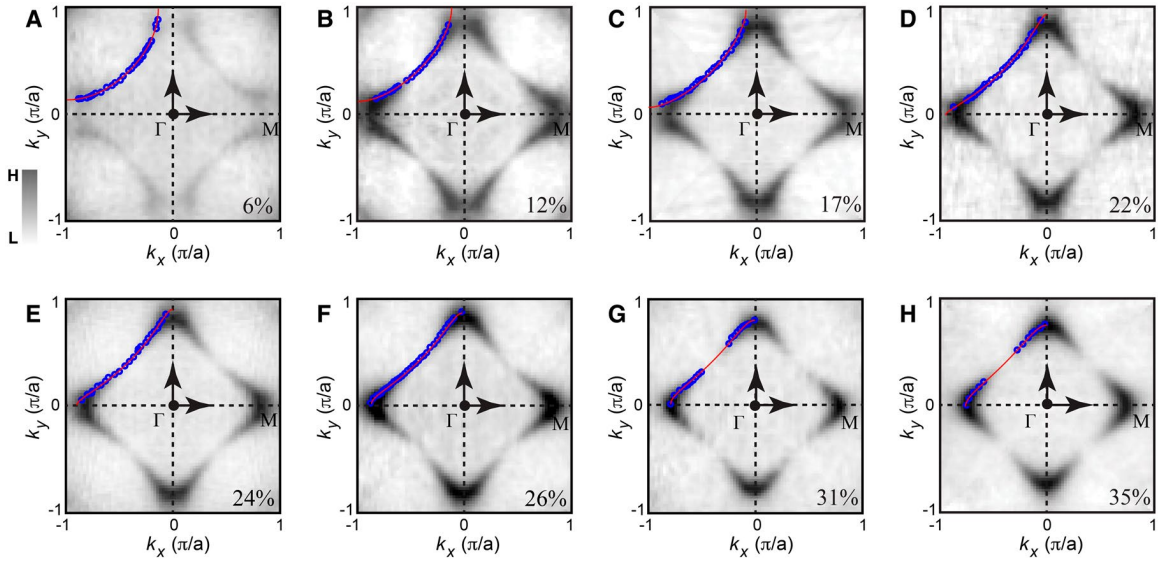

*Figure 6: Lifshitz transition in high-$T_C$ cuprate LSCO. (A-H) Doping-dependent in-plane Fermi surface maps of LSCO thin films. All Fermi surfaces are fourfold symmetrized in the first BZ. Fermi momenta $k_F$ (blue circles) are derived from the MDC analysis. Red curves are the tight-binding simulations at 70 eV.*

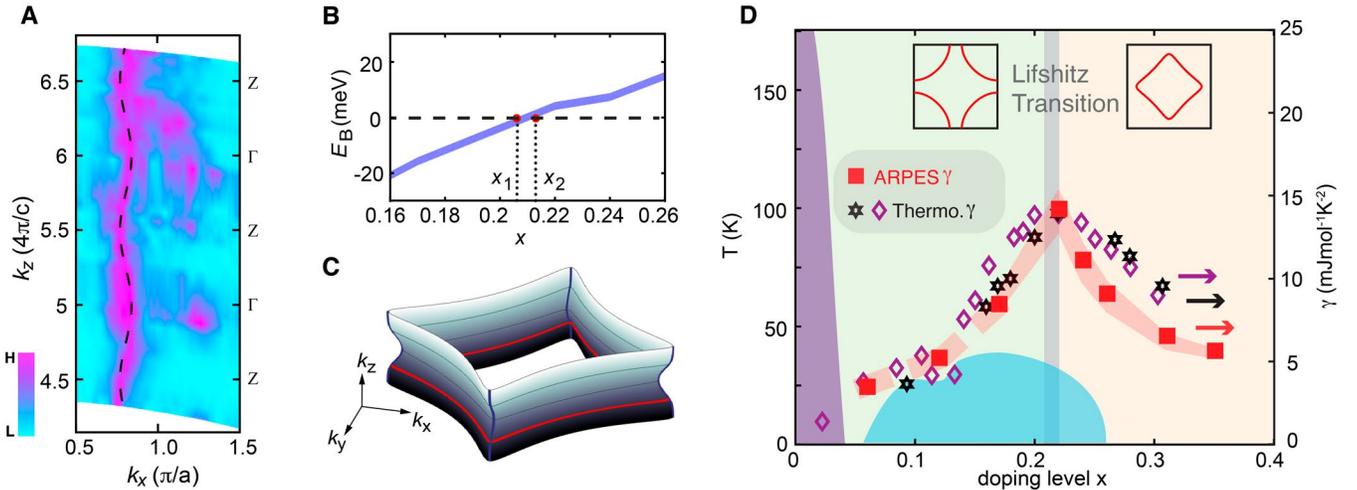

*Figure 7: The role of Lifshitz transition on the specific heat anomaly in LSCO. (A) $k_z$-dependence of Fermi surface map along antinodal direction at $x = 0.22$ sample. The three-dimensionality of Fermi surface structure broadens the Lifshitz transition point to a finite doping range. (B) Doping-dependent binding energy near the antinode. $x_1$ and $x_2$ are the lower and upper bounds of the transition range. (C) Three-dimensional Fermi surface illustration for $x = 0.22$ sample. The red contour is the in-plane Fermi surface collected at 70 eV. (D) Doping-dependent electronic specific heat in LSCO. Gray vertical area represents the transition range. The green and orange areas indicate the hole-like and electron-like Fermi surface topologies, respectively.*

The strength of *in-situ* MBE-ARPES goes beyond the case studies presented in this paper. One unique capability of MBE-grown thin films is the ability to fabricate heterostructures with tailored properties. Such a capability can be further enhanced by combining oxide MBE with other materials synthesis capabilities, such as chalcogenide MBE and 2D materials fabrications. This is exemplified by the FeSe films grown on various oxide substrates and buffer layers for interface engineering

[41–43]. In addition, once the basic electronic structure of the thin films is well characterized at the ARPES beamline, samples can be further studied by other synchrotron techniques, such as soft X-ray ARPES [44] and resonant inelastic X-ray scattering (RIXS), as well as using free electron lasers [45]. These examples showcase the enormous potential of such an approach in the field of condensed matter physics, calling for further development of experimental capabilities where ad-

vanced sample fabrication and *in-situ* characterization are coupled, particularly at synchrotron facilities.

## Acknowledgment

We thank Z.-X. Shen, R. G. Moore, Z. Chen, B. Y. Wang, and R. Wang for stimulating discussions and collaborations. We also thank R. Reininger for the conceptual optical design of the PGM branchline, and the SSRL Beam Line System team led by T. Rabedeau for the construction of the beamline. Stanford Synchrotron Radiation Lightsource, SLAC National Accelerator Laboratory, is supported by the U.S. Department of Energy, Office of Science, Office of Basic Energy Sciences under Contract No. DE-AC02-76SF00515. We acknowledge the support of the U.S. Department of Energy, Office of Science, Office of Basic Energy Sciences, Division of Material Sciences and Engineering, under Contract No. DE-AC02-76SF00515.

## Disclosure statement

No potential competing interest was reported by the authors. ∎

## References

1. M. Imada, A. Fujimori, and Y. Tokura, *Rev. Mod. Phys.* **70** (4), 1039 (1998).
2. H. Y. Hwang et al., *Nature Mater.* **11** (2), 103 (2012).
3. A. Damascelli, Z. Hussain, and Z. X. Shen, *Rev. Mod. Phys.* **75** (2), 473 (2003).
4. J. A. Sobota, Y. He, and Z.-X. Shen, *Rev. Mod. Phys.* **93** (2), 025006 (2021).
5. E. J. Monkman et al., *Nature Mater.* **11** (10), 855 (2012).
6. H. C. Xu et al., *J. Electron Spectrosc. Relat. Phenom.* **200**, 347 (2015).
7. C. K. Kim et al., *J Electron Spectrosc.* **257**, 146775 (2022).
8. K. Horiba et al., *Rev Sci Instrum.* **74** (7), 3406 (2003).
9. Z. Chen et al., *Science.* **373** (6560), 1235 (2021).
10. Y. Zhong et al., PNAS, **119**, e2204630119 (2022).
11. Stanford Synchrotron Ladiation Lightsource (SSRL), https://www.ssrl.slac.stanford.edu/.
12. Beamline 5-2, SSRL, https://www.ssrl.slac.stanford.edu/content/beam-lines/bl5-2.
13. Beamline 5-4, SSRL, https://www.ssrl.slac.stanford.edu/content/beam-lines/bl5-4.
14. S.-D. Chen et al., *Nature* **601** (7894), 562 (2022).
15. S.-D. Chen et al., *Science* **366** (6469), 1099 (2019).
16. W. S. Lee et al., *J Phys-Condens Mat* **21** (16), 164217 (2009).
17. M. Hashimoto et al., *Nature Phys.* **10** (7), 483 (2014).
18. J. J. Lee et al., *Nature* **515** (7526), 245 (2014).
19. Y. Zhang et al., *Phys. Rev. Lett* **117** (11), 117001 (2016).
20. Y. Zhang et al., *Phys. Rev. B* **94** (11), 115153 (2016).
21. H. Pfau et al., *Phys. Rev. Lett* **123** (6), 066402 (2019).
22. H. Pfau et al., *Phys. Rev. B* **99** (3), 035118 (2019).
23. H. Pfau et al., *Phys. Rev. B* **103** (16), 165136 (2021).
24. PEAK, Scienta Omicron, https://scientaomicron.com/en/Instruments/Electron-Analysers/PEAK.
25. E. H. Lieb and F.-Y. Wu, *Phys. Rev. Lett.* **20** (25), 1445 (1968).
26. B. J. Kim et al., *Nat Phys* **2**, 387 (2006).
27. M. Kohno, ", *Phys. Rev. Lett* **105** (10), 106402 (2010).
28. K. M. Shen et al., *Phys. Rev. Lett* **93** (26), 267002 (2004).
29. A. Lanzara et al., *Nature* **412** (6846), 510 (2001).
30. N. Doiron-Leyraud et al., *Nat Commun* **8** (1), 2044 (2017).
31. C. Proust and L. Taillefer, *Annu. Rev. Condens. Matter Phys.* **10** (1), 409 (2019).
32. I. Božović et al., *Nature* **536** (7616), 309 (2016).
33. C. C. Tsuei et al., *Phys. Rev. Lett.* **65** (21), 2724 (1990).
34. W. Wu et al., *Phys. Rev. X* **8** (2), 021048 (2018).
35. T. Yoshida et al., *Phys. Rev. B* **74** (22), 224510 (2006).
36. V. B. Zabolotnyy et al., *Phys. Rev. B* **76** (6), 064519 (2007).
37. T. Kondo et al., *J Electron Spectrosc* **137-140**, 663 (2004).
38. I. K. Drozdov et al., *Nat Commun* **9** (1), 5210 (2018).
39. N. Momono et al., *J. Phys. Soc. Jpn.* **71** (12), 2832 (2002).
40. C. Girod et al., *Phys. Rev. B* **103** (21), 214506 (2021).
41. S. N. Rebec et al., *Phys. Rev. Lett* **118** (6), 067002 (2017).
42. S. N. Rebec et al., *PNAS*, **116** (34), 16687 (2019).
43. T. Jia et al., *Adv. Sci.* **8** (9), 2003454 (2021).
44. Z. Chen et al., *Matter* **5** (6), 1806 (2022).
45. S. Gerber et al., *Science* **357** (6346), 71 (2017).